\documentclass{amsart}
\newtheorem{theorem}{Theorem}
\begin{document}
\sloppypar
\author{Ilja Schmelzer}
\title{Space-geometric interpretation of standard model fermions}

\begin{abstract}
Based on the geometric interpretation of the Dirac equation as an evolution equation on the three-dimensional exterior bundle $\Lambda(\mathbb{R}^3)$, we propose the bundle $(T\otimes\Lambda\otimes\Lambda)(\mathbb{R}^3)$ as a geometric interpretation of all standard model fermions.  The generalization to curved background requires an ADM decomposition $M^4\cong M^3\times\mathbb{R}$ and gives the bundle $(T\otimes\Lambda\otimes\Lambda)(M^3)$.  As a consequence of the geometric character of the bundle there is no necessity to introduce a tetrad or triad formalism.  Our space-geometric interpretation associates colors as well as fermion generations with directions in space, electromagnetic charge with the degree of a differential form, and weak interactions with the Hodge $*$ operator.

The space-geometric interpretation leads to different physical predictions about the connection of SM with gravity, but gives no such differences on Minkowski background.

\end{abstract}

\maketitle

\section{Introduction}

There is a surprisingly large number of ocurrences of the number three in the standard model: Three colors, three fermion generations, $2^3$ fermions in each generation, $2^3$ components of a single Dirac fermion. The main idea of our approach is to associate these occurrences of three with the three spatial dimensions.

This idea is in obvious conflict with the relativistic unification of space and time into an indivisible four-dimensional spacetime manifold.  But even in the context of general relativity the four-dimensional spacetime interpretation is not the only possible interpretation of the gravitational field.  An alternative geometric interpretation of the gravitational field is known as ADM decomposition \cite{ADM}, sometimes also called geometrodynamics.  It depends on a folitation of spacetime $M^4\cong M^3\times\mathbb{R}$ defined by a time coordinate $t$.  For a given foliation, the four-dimensional spacetime metric $g_{\mu\nu}$ decomposes into three three-dimensional objects: A positive-definite spatial metric, a three-dimensional vector field, and a positive scalar field.  This decomposition is widely used in the canonical quantum gravity program.  From point of view of the relativistic spacetime concept it is acknowledged only as an intermediate technical tool, which allows to define a Hamilton formalism for general relativity --- a prerequisite of the canonical quantization program.  The widely expressed hope is that after successful quantization the resulting theory no longer depends on this non-relativistic tool.

The author of this paper, instead, believes that this three-dimensional geometric interpretation is more fundamental than the spacetime interpretation.  In this case, there should be a similar three-dimensional geometric interpretation for matter fields too. The general property of this type of interpretation is that all geometric objects are three-dimensional.  For a given ADM decomposition $M^4\cong M^3\times\mathbb{R}$ they should be objects on the spatial manifold $M^3$.  The time $t$ is only a parameter of dynamics without any geometric interpretation. This type of geometric interpretation we will call here {\em space-geometric\/}.

Our space-geometric interpretation of standard model fermions is based on a space-geometric interpretation of a single fermion as a section of the exterior bundle $\Lambda(M^3)$.  The four-dimensional Dirac equation becomes an evolution equation on this bundle.  An equivalent interpretation has been independently found by Daviau \cite{Daviau}.  It is not the purpose of this paper to give an overview of the research for geometric interpretations of fermions, which can be found, for example, in \cite{Fauser}.  But it seems necessary to discuss the relation between standard spinors and space-geometric fermions in detail.  We show that our space-geometric interpretation of fermions is equivalent to standard theory in the Minkowski case.  But it gives a different connection between  fermions and the gravitational field which is in principle observable in a graviational analogon of Bohm-Aharonov experiments.

Then we propose to interpret all standard model fermions as sections of the bundle $(T\otimes\Lambda\otimes\Lambda)(M^3)$.  Here we identify the three generations of fermions with a vector index and the octet of fermions of one generation with another factor $\Lambda(\mathbb{R}^3)$.  This gives not only the correct number of eight fermions, but also a graduation $(1,3,3,1)$ by the degree of a differential form --- as necessary to  describe two leptons and two quark flavours with three colors for each quark. The actions of gauge fields on the octet are in a natural way associated with geometric operators.

\section{Space-geometric representation of spinors} \label{Daviau}

The classical Dirac algebra is defined by the operators $\gamma^i$ with commutation relations $\{\gamma^i,\gamma^j\}=\eta^{ij}$ --- the Clifford algebra $Cl(1,3)$.  But there are also some other operators which play an important role in the standard model: The complex structure, which is defined by the operator $i$, and a real structure defined by the operator $C$, which is necessary to define Majorana mass terms.  The commutation relations of this ``extended Dirac algebra'' $\{\gamma^\mu, i, C\}$ are: $\{\gamma^\mu,\gamma^\nu\}=2\eta^{\mu\nu}$, $[i,\gamma^\mu]=0$, $[C,\gamma^\mu]=0$, $\{C,i\}=0$, $C^2=1$, $i^2=-1$.

The three-dimensional or space-geometric representation of spinors is based on the following isomorphism:

\begin{theorem}\label{t1}
The ``extended Dirac algebra'' $\{\gamma^\mu, i, C\}$ is isomorph to the Clifford algebra $Cl(3,3)$.   The Clifford algebra generators are the $\gamma^i$ for $i>0$ and the three operators  $\beta^i$ defined by

\begin{eqnarray*}
\beta^1&=&C,\\
\beta^2&=&Ci,\\
\beta^3&=&\gamma^5=-i\gamma^0\gamma^1\gamma^2\gamma^3.
\end{eqnarray*}
\end{theorem}

Proof: An explicit check proves that the operators $\gamma^i,\beta^i$ fulfil the properties required for the generators of $Cl(3,3)$ (they anticommute, $(\beta^i)^2=-(\gamma^i)^2=1$).  We have already expressed the generators of $Cl(3,3)$ in terms of the extended Dirac algebra $\{\gamma^\mu, i, C\}$.  The reverse expressions can be easily found:

\begin{eqnarray}
C &=& \beta^1\\
i &=& \beta^1\beta^2\\
\gamma^0 &=& \beta^1\gamma^1\beta^2\gamma^2\beta^3\gamma^3
\end{eqnarray}

It remains to prove that there are no other relations in  $\{\gamma^\mu, i, C\}$.  This can be done by an explicit computation of the dimension of the extended Dirac algebra.  We have six generators  $\{\gamma^\mu, i, C\}$, for each pair of generators we have one commutation resp. anticommutation rule, the square of all generators is $\pm 1$, and there are no other independent relations.  Therefore an arbitrary monom may be transformed into a monom where every generator appears at most once.  There are $2^6$ such monoms, which gives the dimension of $Cl(3,3)$. qed.

The Clifford algebra $Cl(3,3)$ acts in a natural way on the three-dimensional Clifford bundle $Cl(\mathbb{R}^3)$.  This bundle is equivalent to the exterior bundle $\Lambda(\mathbb{R}^3)$ which consists of inhomogeneous differential forms $\varphi=\varphi_\kappa(x) (dx^1)^{\kappa_1}(dx^2)^{\kappa_2}(dx^3)^{\kappa_3}$, $\kappa=(\kappa_1,\kappa_2,\kappa_3)\in(\mathbb{Z}_2)^3$.  The operators $\gamma^i$ act as in standard Hodge theory and define the Hodge theory Dirac operator

\begin{equation}
\label{D3} D_3 = i \gamma^i \partial_i
\end{equation}

(see, for example, \cite{Pete}).  The action of the operators $\beta^i$ follows from the following formula:

\begin{equation}
\label{betadef} (\beta^i\gamma^i\varphi)_\kappa(x) = (-1)^{\kappa_i} \varphi_\kappa(x)
\end{equation}

It is not hard to see the following:

\begin{theorem}\label{t2}
The representation of the extended Dirac algebra $\{\gamma^\mu, i, C\}$ on $\Lambda(\mathbb{R}^3)$ defined by (\ref{D3}), (\ref{betadef}) and theorem \ref{t1} is equivalent to the standard representation.
\end{theorem}

Theorem \ref{t1} gives now the following four-dimensional Dirac operator:

\begin{equation}
\label{D4} D_4 = i\gamma^0 \partial_0 + D_3
\end{equation}

The definition of the Lagrange formalism simplifies if we use the following variant of the Dirac operator:

\begin{equation}
\label{D} D = -i\gamma^1\gamma^2\gamma^3 D_4 = \iota \partial_t + \iota \alpha^i \partial_i
\end{equation}

where  $\iota=\beta^1\beta^2\beta^3$, $\iota^2=-1$, $\{\iota,\gamma^i\}=0$, $[\iota,\alpha^i]=0$.  The equivalence is obvious: $D\varphi=0$ iff $D_4\varphi=0$.  Now we can use the standard Euclidean scalar product on $\Lambda(\mathbb{R}^3)$ to define the Langrange formalism as

\begin{equation}
\label{Lcontinuous}L = (\varphi D \varphi)
\end{equation}

The Dirac operator considered here should not be mingled with the four-dimensional Dirac-K\"ahler operator defined on the spacetime bundle $\mathbb{C}\otimes\Lambda(\mathbb{R}^4)$ which has real dimension 32.  It defines an evolution equation on the spatial bundle $\Lambda(\mathbb{R}^3)$ with real dimension 8.  Time is only a parameter of dynamics and has no geometric interpretation. Manifest Lorentz invariance is lost.  Only the equivalence theorem proves that Lorentz invariance will be recovered for observable effects.

This representation also violates manifest symmetry in space:  The complex structure $i=\beta^1\beta^2$ as well as the operator $\gamma^5=\beta^3$ depend on a preferred direction in space.  The introduction of three generations of fermions associated with the three spatial directions may be used to get rid of this asymmetry.

A similar "space Clifford" representation of the Dirac equation has been considered by Daviau \cite{Daviau}, see also \cite{Fauser}.

\subsection{Discussion of the differences between geometric fermions and standard spinors}

In section \ref{Daviau} we have found the equivalence between the standard Dirac equation and our geometric representation of the Dirac equation as an evolution equation on the bundle $\Lambda(\mathbb{R}^3)$.  On the other hand, the action of the Lorentz group $SO(1,3)$ on fermions is completely different.  Indeed, for the standard Dirac equation we have a spinor representation of the Lorentz group, defined by the generators

\begin{eqnarray}
\sigma_{ij} &=& \gamma^i\gamma^j\\
\label{alpha} \alpha^i    &=&  \frac{1}{2}\iota\epsilon_{ijk}\sigma_{jk} = \gamma^0\gamma^i.
\end{eqnarray}

Instead, any geometric interpretation uniquely defines a representation of $SO(3)$ which therefore cannot be a spinor representation.  Any diffeomorphism of the underlying manifold uniquely defines the related transformation of any geometric (tensor) bundle. This holds also for the bundle $\Lambda(\mathbb{R}^3)$ of our geometric interpretation.  Indeed, the representation of the group of rotations $SO(3)$ on $\Lambda(\mathbb{R}^3)$ is generated by

\begin{equation}
(\sigma')_{ij}=\gamma^i\gamma^j-\beta^i\beta^j
\end{equation}

and gives a non-spinor representation of the group $SO(3)$.  The extension to the Lorentz group $SO(3,1)$ can be obtained in analogy to (\ref{alpha}):

\begin{equation}
(\alpha')^i=\frac{1}{2}\iota\epsilon_{ijk}(\sigma')_{jk}.
\end{equation}

There seems to be a contradiction:  On one hand, the equivalence theorems \ref{t1}, \ref{t2} seem to prove that there is no physical difference on the Minkowski background.  On the other hand, we have different actions of the Lorentz group --- and these actions are different already on the Minkowski background.  It sounds strange that this difference should not have any physical consequences.  Nonetheless, the following consideration shows that there is no contradiction:

First, it is important to remember that a complete, correct definition of a physical theory can be given in a single system of coordinates.  Such a definition can have the form: ``There exists such a system of coordinates $X$ so that $\ldots$''.  This particular system of coordinates allows to define everything --- especially every observer.  This holds for all physical theories --- theories with preferred background or preferred frames as well as for covariant theories.  For example, the Lorentz symmetry of the Maxwell equations has been found much later than the theory itself.  Nonetheless, it would be nonsensical to claim that Maxwell's definition of EM theory was incomplete.  In a similar way, we do not need the Lorentz transformations to define Dirac theory completely.

What is, when, the role of Lorentz transformations?  Such symmetry transformations allow us to prove that the system of coordinates used in this definition cannot be uniquely detected by observation.  This is an interesting and important physical {\em property\/}, but in no way necessary for a correct {\em definition\/} of the theory.   Now, if there are two versions of the action of the Lorentz group, above versions may be used to prove that the preferred frame remains unobservable.  The result remains valid --- the preferred frame unobservable --- independent on metaphysical speculations which of the two versions defines the ``true action'' of the Lorentz group.

There is also another point which is worth to be discussed --- the well-known spin-statistics theorem.  According to this theorem, particles with half-integer spin should be quantized as fermions, particles with integer spin as bosons.   Now, in standard theory particles with half-integer spin have spinor representations of the Lorentz group, and particles with integer spin have tensor representations.  What about our geometric representation?  Maybe because we have a tensor representation of the group of rotations, geometric fermions have to be quantized as bosons?   In this case, geometric fermions would be ruled out.   Fortunately, this does not happen.  The properties used in the usual proofs of the spin-statistics theorem are operator properties, and these operator properties do not depend on the representation of the Lorentz group.

Nonetheless, there is a domain where the difference becomes important --- gravity.   The difference between the Minkowski case and a curved background is that in the Minkowski case we have no way to compare, by observation, an unrotated spinor field with a rotated one.  Therefore we have no physical effects which depend on the action of the Lorentz group.  Instead, let's consider on a curved background an experiment of Bohm-Aharonov type.  The connection defines a nontrivial element of $SO(3,1)$ for every a closed path.  Then we have to use the representation of the Lorentz group to compute a related phase transformation for the inner space.  This transformation gives nontrivial interference effects.  Now, these observable interference effects already depend on the representation of the Lorentz group.   Thus, a physical difference between spinor representations and geometric representations exists.  But it becomes physical only on curved background, in the domain of semiclassical quantum gravity.

Possible ways to generalize the Dirac equation to curved background we consider below in sec. \ref{curved}.

\section{Geometric representation of the standard model fermions}

Based on this equation for a single fermion we propose here a similar space-geometric interpretation for the whole fermionic part of the standard model: The bundle $(T\otimes\Lambda\otimes\Lambda)(\mathbb{R}^3)$.  Here the first vector index defines the generation, the second factor defines the eight fermions in a given generation, and the last factor defines the fermion itself.

In this interpretation we have not only left-handed but also right-handed neutrinos, in agreement with most modern versions of the standard model which handle massive neutrinos. The association of the exterior bundle with an octet seems quite natural: We have a nice identification of the colors of the quarks with directions in space. Electromagnetic charge may be associated with the degree of the related differential form, and the Hodge star operator connects pairs of fermions which are doublets of weak interaction.

\subsection{Three generations allow an isotropic complex structure} \label{complex}

The complex structure $i=\beta^1\beta^2$ on a single space-geometric fermion depends on a preferred direction in space.  This holds as well for the whole octet
$(\Lambda\otimes\Lambda)(\mathbb{R}^3)$.  But the introduction of a vector index for the three fermion generations allows to heal this asymmetry:  It allows to define a direction-independent complex structure.   The receipt is simple:  On the fermion generation which is defined by spatial index $i$ we define the complex structure as $i=\beta^j\beta^k$, $j=i+1 \mbox{ mod } 3, k=i+2 \mbox{ mod } 3$.

Thus, if we introduce three generations of fermions, the complex structure becomes isotropic, we do not have to prefer a direction.  The asymmetry in space for a single space-geometric fermion, initially a strong symmetry argument against space-geometric fermions, becomes now an advantage --- it allows to explain the three generations of fermions which we observe in nature.  Note that this symmetrization requires the association of the three fermion generation with directions in space --- $(T\otimes\Lambda\otimes\Lambda)(\mathbb{R}^3)$ instead of $(\Lambda\otimes\Lambda)^3(\mathbb{R}^3)$.

Of course, if we include mass terms into the considerations, isotropy in space has to be broken if we associate fermion generations with directions in space.  But this seems unproblematic if we interpret, as usual, masses as a side effect of spontaneous symmetry breaking.  A direction-dependent complex structure would be a much more serious, more fundamental violation of spatial symmetry.

\subsection{Interaction terms with Standard Model gauge fields}

Let's consider now the question how the gauge field actions of the standard model may be defined in a natural way on the bundle $(T\otimes\Lambda\otimes\Lambda)(M^3)$.

The Clifford algebra $Cl(3,3)\cong M_8(\mathbb{R})$ acts on the ``inner space'' of degrees of freedom of a single bispinor as well as on the ``octet space''.  Let's use the denotations $\beta^i, \gamma^i$ for the action on the inner space and $\beta_i, \gamma_i$ for the action on the octet space, and introduce the following special abbreviations:  $\gamma_0=\beta_1\gamma_1\beta_2\gamma_2\beta_3\gamma_3$,  $\beta=\beta_1\beta_2\beta_3$, $\gamma=\gamma_1\gamma_2\gamma_3$.  Note that we already have defined the operators $i=\beta^1\beta^2$, $\gamma^5=\beta^3$ on the inner space.  The resulting set of operators allows to define arbitrary gauge actions on the octet.

The interesting question is if the gauge actions which appear in the standard model are, in whatever sense, more natural than other possible types of gauge actions on the bundle $(T\otimes\Lambda\otimes\Lambda)(\mathbb{R}^3)$.  In this case, our geometric model would have some explanatory power.  Now, the standard model gauge actions are in very nice correspondence with natural actions on this bundle.

For example, for the actions of the photon $\gamma_{EM}$, the $Y$- and the $W$-boson of electroweak theory such natural candidates are:

\begin{equation}
\gamma_{EM}= \frac{i}{6}\sigma - \frac{i}{2}\gamma_0;  \hspace{0.5cm}
Y = \frac{i}{6}\sigma - \frac{i}{4}(1-\gamma^5)\gamma_0;
\end{equation}
\begin{equation}
W_1= \frac{-1}{2}(1+\gamma^5)\beta;  \hspace{0.5cm}
W_2= \frac{i}{2}(1+\gamma^5)\gamma;  \hspace{0.5cm}
W_3= \frac{i}{2}(1+\gamma^5)\gamma_0.
\end{equation}

with $\sigma=\beta_1\gamma_1\beta_2\gamma_2+\beta_2\gamma_2\beta_3\gamma_3+\beta_3\gamma_3\beta_1\gamma_1$. Electromagnetic charge would be associated with the degree of the form in $\Lambda(\mathbb{R}^3)$.  The operator $\gamma$ used to define $W_2$ is the well-known Hodge $*$ operator, $\gamma_0$ used to define $W_3$ defines the graduation in $\Lambda(\mathbb{R}^3)$.
Strong interaction may be defined by the following rules:

\begin{itemize}

\item On the inner space, strong interaction is a vector gauge action;

\item Strong interaction does not change the degree;

\item All elements commute with $\beta$, $\gamma$ and $\sigma$.

\item The center of the remaining algebra has to be excluded.

\end{itemize}

Note that because of the use of $i$ in these gauge actions they become spatially symmetric only if we consider all three generations, as discussed in sec \ref{complex}.

Especially the subgroup $SO(3)\subset SU(3)$ is associated with the group of rotations. This set of rules looks sufficiently simple.  On the other hand, some properties remain unexplained, especially the chiral character of weak interaction.

\section{Generalization to curved background}\label{curved}

We have already noted that generalization to curved background gives different physical predictions for standard spinors (tetrad formalism) and geometric fermions, because of the different representations of the Lorentz group which are involved.  Let's consider now shortly some qualitative features of the generalization of the space-geometric interpretation to curved background.

To define an appropriate bundle which generalizes  $(T\otimes\Lambda\otimes\Lambda)(\mathbb{R}^3)$ to curved background we need an ADM decomposition --- a foliation of the spacetime manifold into space and time $M^4\cong M^3\times\mathbb{R}$ defined by some global time coordinate $t$.  This defines a decomposition of the four-dimensional spacetime metric $g_{\mu\nu}$ into three three-dimensional geometric objects --- a positive scalar function $\rho$, a three-vector field $v^i$ and a positive definite three-metric $\sigma^{ij}$ --- in the following way:

\begin{eqnarray}
\nonumber  g^{00} \sqrt{-g} &=& \rho\\
\label{g}  g^{0i} \sqrt{-g} &=& \rho v^i\\
\nonumber  g^{ij} \sqrt{-g} &=& \rho v^i v^j - \sigma^{ij}
\end{eqnarray}

On such a decomposition we have now a natural generalization of the bundle $(T\otimes\Lambda\otimes\Lambda)(\mathbb{R}^3)$ --- the bundle $(T\otimes\Lambda\otimes\Lambda)(M^3)$.

A more subtle question is if there exists a natural generalization of the Dirac operator on the bundle $(T\otimes\Lambda\otimes\Lambda)(M^3)$.  Here we may remember an idea proposed by Hestenes \cite{Hestenes} --- the identification of electroweak doublets with the pair of fermions described by the four-dimensional external bundle $\Lambda(M^4)$.  This identification defines a natural Dirac-K\"ahler operator on these doublets.  It remains to identify the exterior bundle $\Lambda(M^4)$ with the electroweak doublets which in our space-geometric interpretation are described as pairs of three-dimensional bundles $\Lambda(M^3)\oplus\Lambda(M^3)$ connected with each other via the Hodge star operator.
In principle, the preferred time coordinate $t$ of the ADM decomposition allows to define a sufficiently natural decomposition $\Lambda(M^4)\cong\Lambda(M^3)\oplus\Lambda(M^3)$.  The first subbundle may be defined as the image of exterior multiplication with $dt$, the second as its orthocomplement.
Details have to be worked out by future research.  It seems not unreasonable to guess that there may be other, more natural candidates for Dirac operators on the bundle $(T\otimes\Lambda\otimes\Lambda)(M^3)$. But already now we can make some qualitative conclusions:

\begin{itemize}

\item We need an ADM decomposition as an additional structure to generalize the space-geometric interpretation to curved background;

\item We do not need a tetrad or triad formalism;

\item The generalization of the space-geometric interpretation to curved background leads to differences in the predictions in the domain of semiclassical quantum gravity in comparison with the classical tetrad approach.   These differences are at least in principle observable.

\end{itemize}

\section{Conclusion}

We have proposed a geometric interpretation of standard model fermions as sections of the bundle $(T\otimes\Lambda\otimes\Lambda)({\mathbb{R}^3})$.   This interpretation may be generalized to curved background in the context of an ADM decomposition $M^4\cong M^3\times \mathbb{R}$ as the three-dimensional geometric bundle $(T\otimes\Lambda\otimes\Lambda)({M^3})$. Time appears only as a scalar parameter of dynamics, without geometric interpretation.
 
The interpretation associates the three colors as well as the three fermion generations with directions in space.   As a consequence,  we obtain different predictions in the domain of interaction with gravity --- especially about phase factors in
experiments of Bohm-Aharonov type with gravitational instead of electromagnetic fields.

We have defined the standard model gauge actions on the bundle $(T\otimes\Lambda\otimes\Lambda)({\mathbb{R}^3})$.  These actions are remarkably simple and natural --- they depend on important geometric properties of and operators on the bundle, like the degree of a differential form, the Hodge $*$ operator, and the group of rotations.

The consideration of curved background suggests that we do not need a tetrad or triad formalism to describe fermions.   This gives an interesting simplicity argument in favour of our interpretation.   What we need is only a foliation of spacetime into space and time --- a quite natural object in a world where space and time seem to be quite different.

\begin{appendix}

\section{Is the preferred frame hypothesis viable?}

A space-geometric interpretation, almost by definition, depends on the choice of the frame or the time coordinate $t$.  Even if there is no difference in the physical predictions, another choice of the frame leads to a different interpretation.

Thus, if the space-geometric interpretation describes some aspect of reality, there should be a preferred frame.  In this context, it seems useful to discuss the viability of the preferred frame hypothesis.   Nonetheless, two remarks are necessary:  First, such a space-geometric representation may be useful as a technical tool even if there is no preferred frame in reality.   Second, we should avoid circular thinking:  The existence of a nice space-geometric interpretation of standard model fermions which we have presented here is a quite strong argument in favour of the preferred frame hypothesis, especially if there is no comparable spacetime interpretation.  It would be nonsensical to reject the preferred frame hypothesis based on the old arguments, without taking this new interpretation into account, and then to reject this interpretation.  Moreover, the interpretation presented here is new, and future research based on it can give additional support for the preferred frame hypothesis.  This has to be taken into account if we start to compare here arguments in favour and against the preferred frame hypothesis.

The most powerful argument against a preferred frame seems to be the success of general relativity --- a theory of gravity where many solutions do not allow a preferred frame. But in \cite{GLET} the author has proposed an theory of gravity with preferred frame which generalizes the Lorentz ether.  In this theory, the Einstein equivalence principle holds, and in some natural limit we obtain the Einstein equations of general relativity.  Moreover, the EEP and the general Lagrangian of the theory are derived from sufficiently natural ether axioms.  This removes empirical as well as important metaphysical arguments against a preferred frame.

On the other hand, there are important arguments in favour of a preferred frame.  First, there is the violation of Bell's inequality.  It excludes Einstein-causal realistic hidden variable theories.  Thus, or we have to give up realism, or we have to give up Einstein causality and to reintroduce a preferred frame for the realistic hidden variables.  Then there is the well-known conflict with the principle of local conservation of energy and momentum in general relativity.  Last not least, there are deep conflicts between relativity and quantum principles known as the ``problem of time'' and ``information loss problem''.

If we evaluate the seriousness of these conflicts between relativity and other physical principles we should be very careful if we want to avoid circular reasoning.  Indeed, these other principles have been questioned and rejected because they are in conflict with the relativistic paradigm --- a very powerful argument against a particular physical principle, but invalid circular reasoning in the context of this discussion.  A detailed consideration would have to evaluate if there is independent evidence against these other principles.  The existence of theories with preferred frame which are compatible with these other principles can prove that there is no such independent evidence.

Such theories exist:  There are hidden variable theories with preferred frame like Bohmian mechanics and Nelson's stochastics.  And there is the ether theory of gravity \cite{GLET} which gives well-defined local energy and momentum densities for the gravitational field.  Given this situation, we conclude that the preferred frame hypothesis is not only viable, but there are sufficiently strong arguments in its favour.

The space-geometric interpretation of standard model fermions is a new, strong argument in favour of a preferred frame.  Now, the existence of geometric interpretations for the observable physical fields become an argument in favour of the preferred frame hypothesis:  There is no comparable spacetime-geometric interpretation of the standard model.   This is an important point, because the geometric intepretation of the gravitational field was one of the most powerful arguments in favour of relativity.

\end{appendix}


\thebibliography{99}

\bibitem{Kaehler} E. K\"ahler, Rendiconti di Matematica (3-4) 21, 425 (1962)

\bibitem{Pete}
G. Pete, {Morse theory}, (1999) http://www.math.u-szeged.hu/gpete/morse.ps

\bibitem{ADM} R. Arnowitt, S. Deser, C.W. Misner, The dynamics of general relativity, in L.Witten (ed.), Gravitation: An introduction to current research, Wiley, NY 1962

\bibitem{Daviau} C. Daviau, Dirac equation in the Clifford algebra of space, in V. Dietrich, K. Habetha, G. Jank (eds.) , proceedings of ``Clifford algebra and their applications in mathematical physics'' Aachen 1996, Kluwer/Dordrecht 1998, 67-87

\bibitem{Fauser} B. Fauser, On the equivalence of Daviau's space Clifford algebraic, Hestenes' and Parra's formulations of (real) Dirac theory, hep-th/9908200

\bibitem{GLET}  I. Schmelzer, A generalization of the Lorentz ether to gravity with general-relativistic limit, gr-qc/0205035

\bibitem{Hestenes} D. Hestenes, Space-time structure of weak and electromagnetic interactions, Found. Phys. vol. 12, 153-168 (1982)
\end{document}